\documentclass[aps,twocolumn,floats,prl,nofootinbib,superscriptaddress,10pt]{revtex4-1}

\usepackage[utf8]{inputenc}

\usepackage{graphicx,amsmath,amsfonts,amssymb,slashed}
\usepackage{bbold,wasysym}

\usepackage{amsmath,amsfonts,bbm,euscript,hyperref}

\usepackage{xcolor}
\usepackage{color}
\usepackage{hyperref}

\hypersetup{colorlinks, citecolor=blue, linkcolor=red, urlcolor=blue}

\usepackage{graphicx,capt-of}

\newcounter{subfloat}

\newcounter{subfloat2}

\newcommand{\nco}{\newcommand}
\nco{\beq}{\begin{equation}} \nco{\eeq}{\end{equation}}
\nco{\beqa}{\begin{eqnarray}} \nco{\eeqa}{\end{eqnarray}}
\def\be{\begin{equation}}
\def\ee{\end{equation}}    
\def\baray{\begin{eqnarray}}
\def\earay{\end{eqnarray}}

\nco{\lra}{\leftrightarrow}

\nco{\sss}{\scriptscriptstyle} \nco{\dphi}{\varphi}
\nco{\lsim}{\mbox{\raisebox{-.6ex}{~$\stackrel{<}{\sim}$~}}}
\nco{\gsim}{\mbox{\raisebox{-.6ex}{~$\stackrel{>}{\sim}$~}}}

\def\IK{\relax{\rm I\kern-.20em K}}
\def\IM{\relax{\rm I\kern-.20em M}}

\def\lsim{\mbox{\raisebox{-.6ex}{~$\stackrel{<}{\sim}$~}}}
\def\gsim{\mbox{\raisebox{-.6ex}{~$\stackrel{>}{\sim}$~}}}
\def\sss{\scriptscriptstyle}

\def\ga{\mathrel{\raise.3ex\hbox{$>$\kern-.75em\lower1ex\hbox{$\sim$}}}}
\def\la{\mathrel{\raise.3ex\hbox{$<$\kern-.75em\lower1ex\hbox{$\sim$}}}}

\def\bp{{\bf p}}
\def\bk{{\bf k}}
\def\bq{{\bf q}}

\begin{document}

\title{Imprints of Primordial Non-Gaussianity on Gravitational Wave Spectrum
}

\author{Caner \"Unal}
\affiliation{ CEICO, Institute of Physics of the Czech Academy of Sciences (CAS), Na Slovance 1999/2, 182 21 Prague, Czechia}
\affiliation{School of Physics and Astronomy, University of Minnesota, Minneapolis, 55455, USA}


\begin{abstract} 
Although Cosmic Microwave Background and Large Scale Structure probe the largest scales of our universe with ever increasing precision, our knowledge about the smaller scales is still very limited other than the bounds on Primordial Black Holes. We show that the statistical properties of the small scale quantum fluctuations can be probed via the stochastic gravitational wave background, which is induced as the scalar modes re-enter the horizon. We found that even if scalar curvature fluctuations have a subdominant non-Gaussian component, these non-Gaussian perturbations can source a dominant portion of the induced GWs. Moreover, the GWs sourced by non-Gaussian scalar fluctuations peaks at a higher frequency and this can result in distinctive observational signatures. We found that the sensitive next-generation-interferometers, which will/could reach $\Omega_{GW}h^2 \sim 10^{-15}$ (such as PTA-SKA, LISA, DECIGO, BBO, CE, ET), can probe $f_{NL} \sim 0.5$ which is even better than the predictions of the next generation CMB experiments. If the induced GW background is detected, but not the signatures arising from the non-Gaussian component, $\zeta = \zeta_G  +  f_{\rm NL} \, \zeta_G^{2}$, this translates into bounds on $f_{\rm NL}$ depending on the amplitude and the width of the GW signal. If the induced GW background is not detected at all, this translates into bounds on scalar fluctuations. The results are independent from the fact that whether PBH are DM or completely negligible part of the current energy density.

\end{abstract}

\maketitle

\paragraph{{\bf Introduction.}}
\vspace{-4mm}

In the inflationary framework, the fluctuations at the largest scales of our observable universe are generated around 50-60 e-folds before the end of the inflationary era. The precise number depends on the details of the forthcoming stages which transfer the energy density from the inflaton to the light degrees of freedom. The  large scale modes, with the corresponding wavenumbers $10^{-4} {\rm Mpc^{-1}} \la k \la  0.1 {\rm Mpc^{-1}}$, are probed by the Cosmic Microwave Background (CMB) and Large Scale Structure (LSS) experiments \cite{Ade:2015lrj,Akrami:2018odb,Dawson:2015wdb}. The spectral distortion experiments can further extend this range up to $10^{4}  {\rm Mpc^{-1}}$  \cite{Fixsen:1996nj,Kogut:2011xw,Andre:2013nfa}. However, even with these improvements, the scales that are probed with precision is only about 8 decades, or around 18 e-folds. Therefore, the remaining $30-40$ e-folds of inflationary era in our horizon is mostly unexplored except the bounds on Primordial Black Holes (PBHs) at various scales. This work aims at obtaining or constraining the physics at scales much smaller than the CMB by studying their imprints on stochastic gravitational waves (GWs). 
 
Density and tensor fluctuations are generated and stretched to super-horizon scales during inflation. In later stages of the cosmic evolution, these modes re-enter the causal horizon and lead to the current structure of our universe. In homogenous and isotropic backgrounds, at linear order, scalar, vector and tensor type fluctuations decouple, but starting from second order in perturbation theory, they couple with each other \cite{Mollerach:2003nq,Ananda:2006af}. We focus on GWs sourced by density perturbations that deviate from exact Gaussianity. We assume density perturbations with a dominant Gaussian (G) piece and a subdominant local type non-Gaussian (NG) piece (except the large $f_{\rm NL}$ section, in which we discuss the case the scalar fluctuations are dominantly NG.). 
    
We further assume an enhancement in scalar curvature power spectrum at scales that re-enter our horizon during radiation dominated era and much smaller than CMB scales. This enhancement can be a result of various reasons.  One reason, for example, could be a feature in inflationary potential (a saddle point or a slope change) \cite{Namjoo:2012aa,Martin:2012pe,Chen:2013aj,Garcia-Bellido:2017mdw,Germani:2017bcs,Motohashi:2017kbs}. Slowing down the inflaton might lead to significant quantum diffusion which influences the statistical properties of the scalar fluctuations \cite{Pattison:2017mbe,Franciolini:2018vbk,Biagetti:2018pjj,Ezquiaga:2018gbw}. Another reason could be the interaction of different fields and efficient particle production during inflation \cite{Anber:2009ua,Bugaev:2013fya,Domcke:2016bkh,Garcia-Bellido:2016dkw,Dimastrogiovanni:2016fuu,Thorne:2017jft,Papageorgiou:2018rfx}; or reduced speed of sound \cite{Langlois:2008qf,Cremonini:2010ua,Achucarro:2012sm,Biagetti:2014asa}. We do not assume any specific model, but a bump with non-vanishing local non-Gaussianity. We found that for a large range of density fluctuations, the resultant induced GWs are detectable at various GW experiments.

The large density fluctuations leading to induced GW background can also produce abundant PBHs or compact massive objects \cite{Saito:2008jc,Bugaev:2010bb}. However, the PBH amount is extremely sensitive to the statistical properties of the density perturbations and threshold for the collapse. Therefore, the results of this work is expected to be valid independent of the fact that PBHs are a significant or a completely negligible fraction of the energy density.

\vspace{1mm}
\paragraph{{\bf Gravitational Waves Induced by Scalar Perturbations During Radiation Domination Era.}}
\label{sec:indgwsrev}

We first give a brief summary of the results for the second order tensor modes sourced by the first order scalar fluctuations  \cite{Mollerach:2003nq,Ananda:2006af,Baumann:2007zm} and continue with the GW background from the non-Gaussian density fluctuations. 

Ignoring vector, first order tensor perturbations and the anisotropic stress (see also \cite{Baumann:2007zm,Weinberg:2003ur,Mangilli:2008bw,Saga:2014jca}), we start with the following metric

\vspace{-4mm}

\begin{equation}
ds^2 = a^2(\eta) \bigg[ - \left( 1+2 \Phi \right) d\eta^2 \, + \bigg\{ (1 - 2 \Phi ) \delta_{i j} +h_{ij} \bigg\} dx^i \, dx^j \bigg] \nonumber
\end{equation}
where $\eta$ is the conformal time, $\Phi$ the scalar gravitational potential and $h_{ij}$ the second order tensor perturbation.

Fourier transformed tensor modes are expressed as
\vspace{-0.4cm}

\begin{equation}
{\hat h}_{ij}(\eta, {\bf x}) = \int \frac{d^3 \bk }{(2 \pi)^{3/2}} \,  {\rm e}^{i \bk \cdot {\bf x}} \,  \sum_{\lambda }  \epsilon^{\lambda}_{ij}(\bk) {\hat h}_{\lambda, \, \bk} (\eta)
\end{equation}

\begin{figure}
\centering{ 
\includegraphics[width=0.1\textwidth]{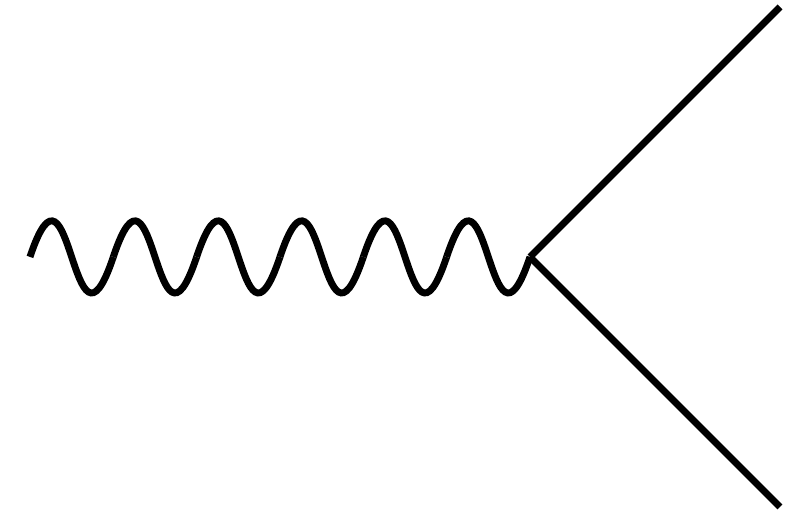}
\hspace{10mm}
\includegraphics[width=0.07\textwidth]{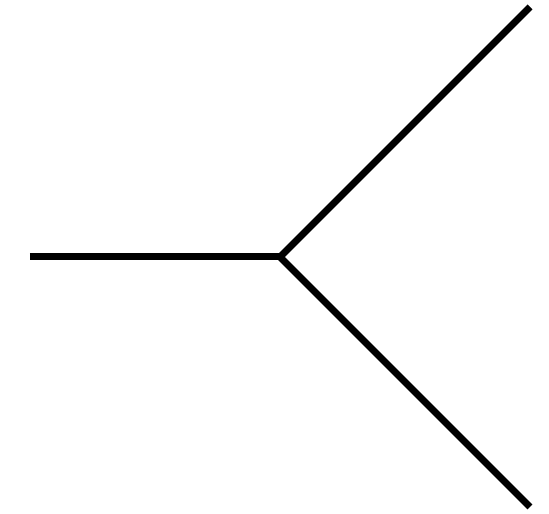}
}
\vspace{-3mm}
\caption{Diagrammatic representation of $h\zeta\zeta$ and $\zeta^3$ vertices
}
\label{figfundvertices}
\end{figure}

The equation of motion for the mode functions reads
\begin{equation}
h''_{\lambda, \, \bk} (\eta) + 2 {\cal H} \, h'_{\lambda, \, \bk} (\eta) + \bk^2  h_{\lambda, \, \bk}(\eta) = 2 {\cal S}_{\lambda, \, \bk} (\eta)
\label{eq:heom}
\end{equation}
where ${\cal S}$ is the source term originating from first order scalar fluctuations (schematically $\zeta+\zeta \to h$, this interaction is shown in the left panel of Fig. \ref{figfundvertices}).
\begin{eqnarray}
 {\cal S}_{\lambda, \, \bk} (\eta) &= &\int \frac{d^3p}{\left(2\pi\right)^{3/2}} \;  \epsilon^{\lambda}_{ij} (\bk) \;  p^i \;  p^j   \Bigg[2 \, \Phi_\bp (\eta)  \Phi_{\bf k-p} (\eta) +  \nonumber\\
&& \!\!\!\!\!\! \!\!\!\!\!\! \!\!\!\!\!\!  \!\!\!\!\!\!   + \left( \Phi_\bp (\eta) + \frac{\Phi'_\bp (\eta) }{{\cal H}(\eta)} \right) \left( \Phi_{\bf k-p} (\eta) + \frac{\Phi'_{\bf k-p} (\eta)}{{\cal H}(\eta)} \right)  \Bigg] 
\end{eqnarray}
where $\Phi_\bk(\eta) = \frac{2}{3} \, T( k \eta) \,  \zeta_\bk$ and $'$ denotes  $\frac{\partial }{\partial \eta}$. Here, $\zeta_\bk$ is  the scalar curvature fluctuation and $T$ is the transfer function given as $T (x) =  \frac{9}{x^2} \left[  \frac{\sin(x/ {\sqrt 3}) }{x/ {\sqrt 3}} -\cos (x/ {\sqrt 3})     \right]$.

The GW power spectrum is defined as usual
\begin{equation}
\left \langle \hat h_{\lambda}( \bk, \eta)  \,\,  \hat h_{\lambda'}( \bk', \eta) \right \rangle \equiv \frac{2\pi^2}{k^3} \delta^3 ( \bk +  \bk') \, \delta_{\lambda \, \lambda'} \,  P_{\lambda} (k) \,,
\end{equation}
Combining all these, we end up with 
\begin{eqnarray}
&& \!\!\! \!\!\! P_{h_i} (\eta, k) \equiv \frac{k^3}{2 \pi^2} \, \sum_{\lambda=\pm} \left \langle h_{i,\lambda} ( {\bf k}) \,  h_{i,\lambda} (  \bk') \right \rangle_{\delta}  \nonumber\\ 
&=&  \frac{64}{81} \frac{1}{ 2\pi^2} \frac{k}{\eta^2} \int \frac{d^3 p \, d^3 q}{(2\pi)^3} \; p^2 \, \sin^2 \theta_{\rm kp} \, U(\bp) \; q^2 \; \sin^2 \theta_{\rm k'q} U(\bq)   \nonumber\\ 
&\times&  \int_0^{\eta} d \eta' \int_0^{\eta} d \eta''   \; \eta' \; \eta'' \;  \sin(k\eta - k\eta') \,  \sin(k' \eta - k' \eta'')     \nonumber\\
&\times &F_T ( p \, \eta', \vert {\bf k-p} \vert\ \eta' )  \, \, F_T ( q \, \eta'', \vert {\bf k'-q} \vert\ \eta'' )  \;  \nonumber\\ 
&\times & \left\langle {\hat \zeta}_{\bf p} \,\, {\hat \zeta}_{\bf k-p}\,\, {\hat \zeta}_{\bf q} \,\, {\hat \zeta}_{\bf k' - q} \right\rangle' \;,
\label{eqPhind}
\end{eqnarray}
where $\cos\theta_{\rm  ab}={\hat a}\cdot{\hat b}$ and $U({\bf \hat n})= \cos(2\phi_{\bf \hat n}) + \sin (2\phi_{\bf \hat n})$. $\langle \rangle_{\delta}$ denotes correlation without $\delta^3 \left( \bk + {\bf k'}\right)$.  $F_T (u,v) = 2 T(u) T(v) + \tilde T(u) \tilde T(v)$,
 where T is defined above and ${\tilde T(x)} = T(x) + x \, \partial_x T(x)$.

The current GW energy density (per logarithmic wavenumber) is linked with the GW power spectrum as
\begin{equation}
\Omega_{\rm GW} (k, \eta_0) h^2 \simeq \frac{\Omega_{{\rm \gamma},0}\, h^2 \,}{12} \frac{ k^2}{\,a^2 H^2 }  \overline{  \sum_\lambda P_{\lambda} } \;, 
\label{eqOGW0}
\end{equation}
where the overbar indicates period averaging, $\Omega_{{\rm \gamma},0}$ is the current radiation density parameter.

We assume the scalar fluctuations which have the form
\begin{equation}
\zeta_\bk = \zeta^G_{ \bk} + f_{\rm NL} \int \frac{d^3p}{(2 \pi)^{3/2}} \zeta^G_{\bp} \, \zeta^G_{ {\bf k-p}} \; ,
\label{eqnscalarfluct}
\end{equation}
then $\zeta$ two-point function becomes 
\begin{equation}
P_\zeta (k) = P^G_\zeta (k) + P^{NG}_\zeta (k)
\label{powerspec}
\end{equation}
and we parametrize the curvature fluctuations as 
\begin{eqnarray}
P^G_\zeta (k) &=& {\cal A} \cdot {\rm exp} \left[ - \frac{\ln^2 \left( k / k_* \right)}{2 \sigma^2} \right] \, \nonumber\\
P^{NG}_\zeta (k) &=& 2 f_{\rm NL}^2 \!\! \int \! \frac{dp}{p} \frac{d \Omega}{4\pi} \frac{k^3 }{\vert \bk- \bp \vert^3} \, P^G_\zeta (p) \, P^G_\zeta (\vert \bk- \bp \vert)  ,  \;\;\;\;\;\;\;
\label{defnG-NG}
\end{eqnarray}
where $k_*$ is some comoving wavenumber that curvature perturbations have a bump at, ${\cal A}$ is the amplitude at that scale and $\sigma$ is the width of the signal in terms of e-folds. (The scalar NG interaction is shown in the right panel in Figure \ref{figfundvertices}). At CMB scales, curvature perturbations are nearly scale invariant and Gaussian, but on small scales the constraints are much weaker. We will focus on the case ${\cal A}$ is much larger than the amplitude of the fluctuations at CMB/LSS scales and $\sigma \sim {\cal O} (1)$.

\begin{figure}
\centering{ 
\includegraphics[width=0.45\textwidth]{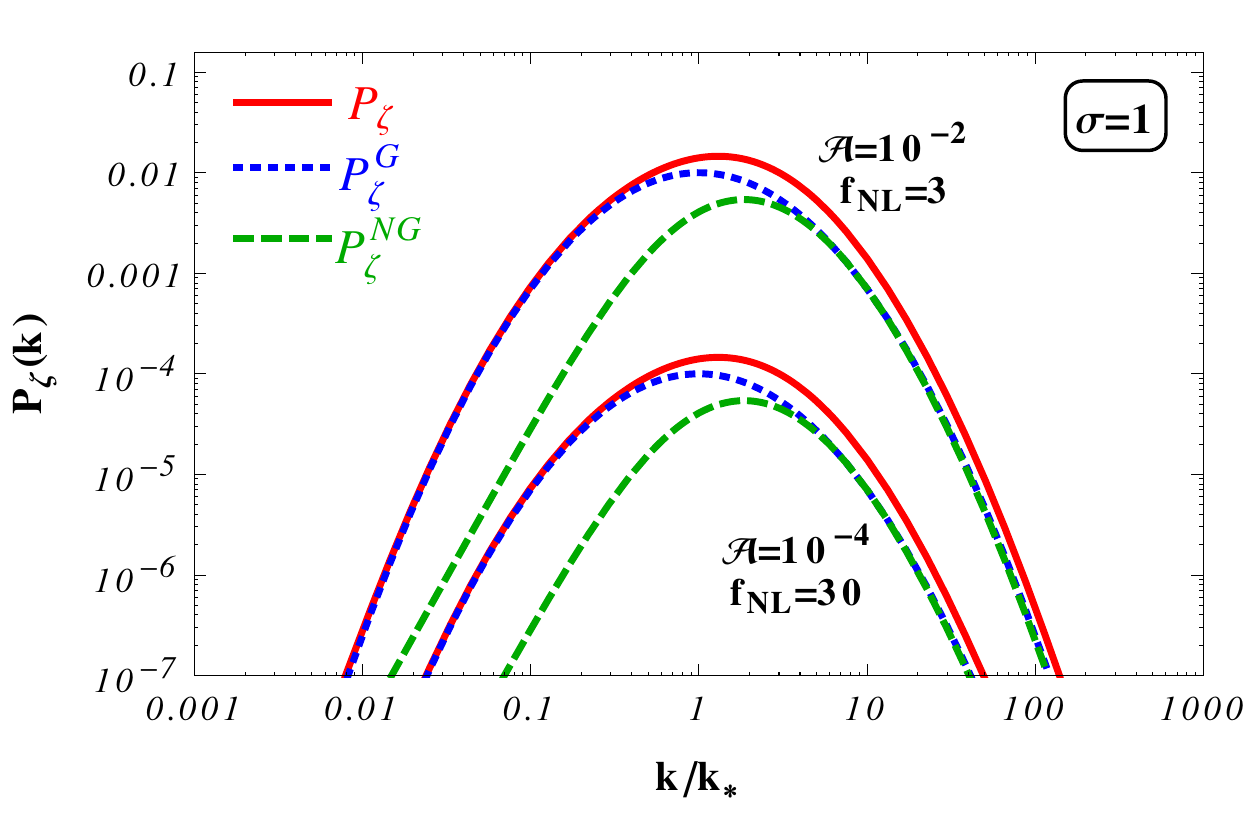}
}
\vspace{-4mm}
\caption{Gaussian and Non-Gaussian components of $P_\zeta$
}
\label{figPZ}
\end{figure}

Figure \ref{figPZ} shows the G and NG components of the scalar power spectrum. As a representative value, we choose ${\cal A}=10^{-2}$, $\sigma=1$ and $f_{\rm NL}=3$. By keeping ${\cal A} \cdot f_{\rm NL}^2$ and $\sigma$ constant, the relative amplitude of G and NG portions stay same (see also the case for ${\cal A}=10^{-4}$, $\sigma=1$ and $f_{\rm NL}=30$). Observe that in both cases, the NG spectrum is subdominant with respect to (or has similar amplitude with) G spectrum at every scale.

\begin{figure}
\centering{ 
\includegraphics[width=0.14\textwidth]{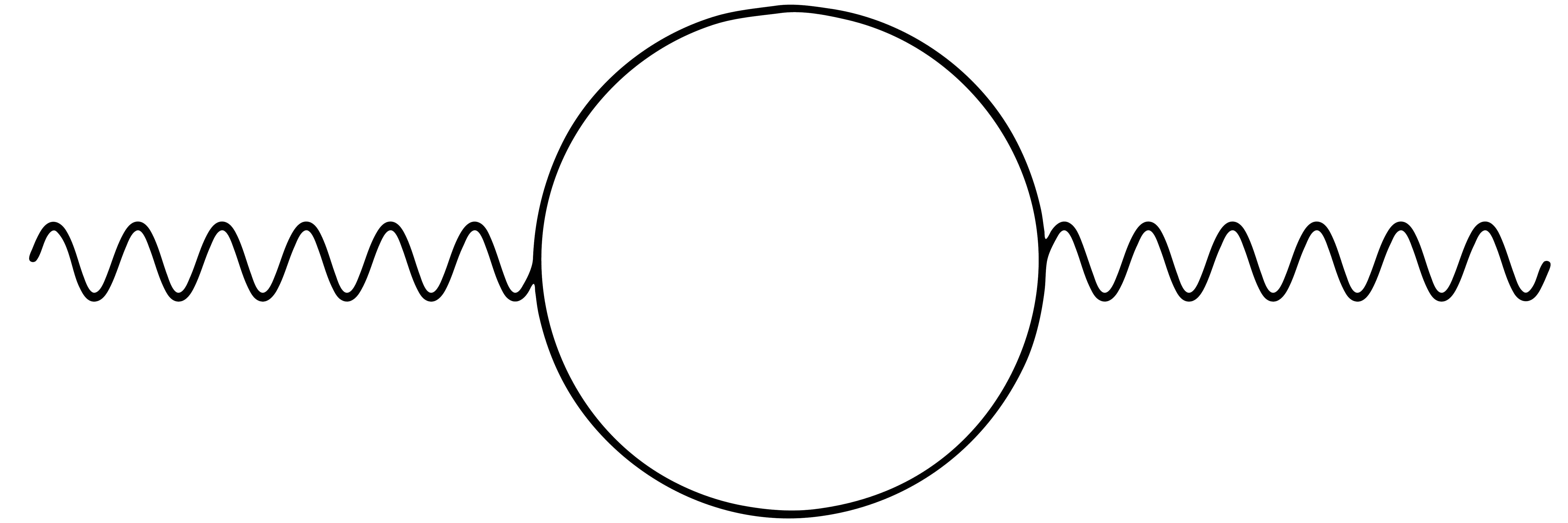}
}
\vspace{-3mm}
\caption{Gaussian Sourcing of GWs
}
\label{figdiagramgauss}
\end{figure}


\paragraph{{\bf The Contribution of the Gaussian Source.}}
When the scalar fluctuations are purely Gaussian, the GWs are sourced via the 1-loop diagram shown in Figure \ref{figdiagramgauss}. A scale-invariant scalar power spectrum produces a scale-invariant GW energy density for the modes that re-enter the horizon during the radiation dominated era as $\Omega_{GW}h^2 \simeq 0.8 \,  \Omega_{\gamma,0} h^2 \, \left(P^{G}_\zeta \right)^2$ \cite{Kohri:2018awv}. For a bumpy scalar spectrum, the resultant GW energy density also becomes bumpy, and the peak of the GW is similar to the  expression above with some ${\cal O}(1)$ correction depending on the width of the scalar signal.



\begin{figure}
\centering{ 
\includegraphics[width=0.2\textwidth]{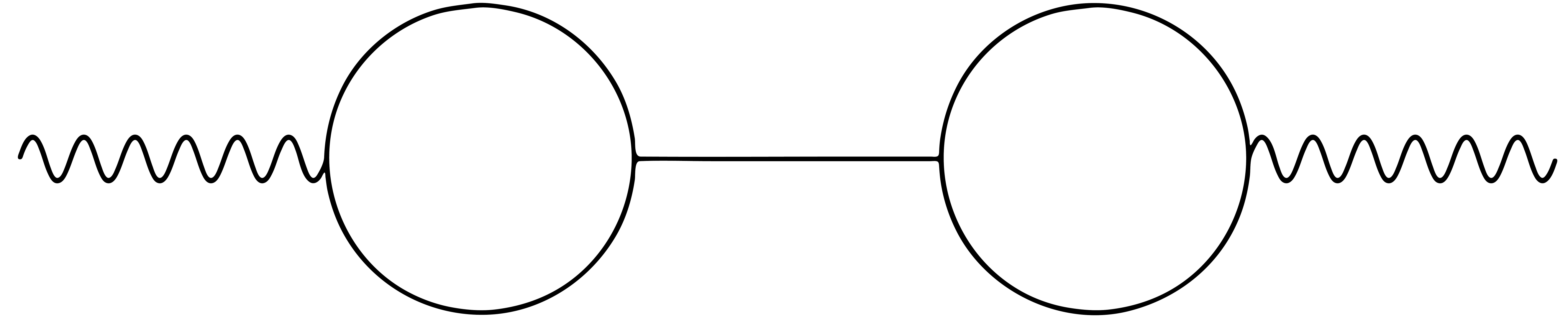}
}
\vspace{-0.3cm}
\caption{Vanishing Non-Gaussian Source due to symmetry
}
\label{figvanishbysymmetry}
\end{figure}


\paragraph{{\bf Contribution of the Non-Gaussian Sources.}}

\begin{figure}
\centering{ 
\includegraphics[width=0.17\textwidth]{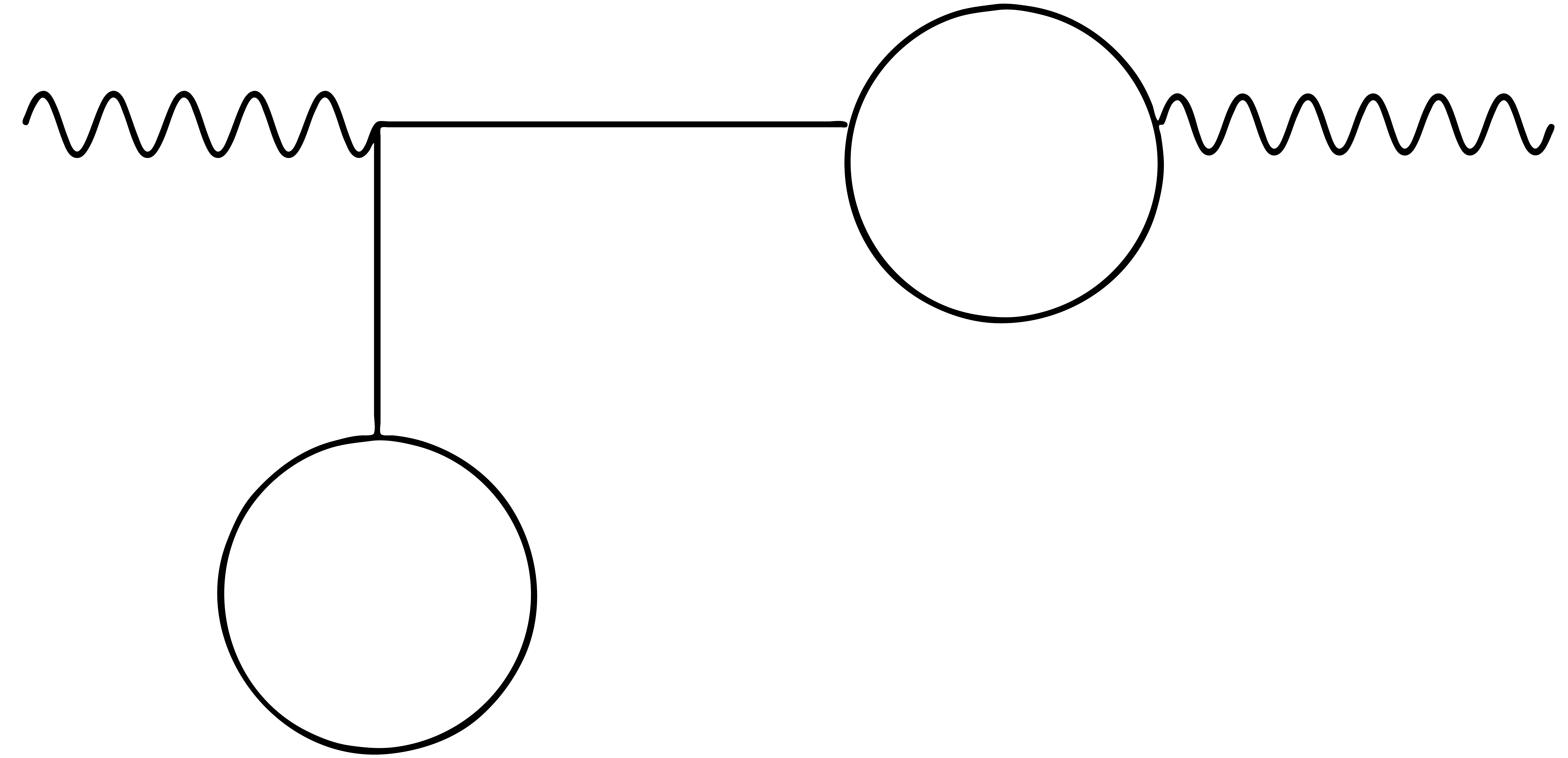} 
\hspace{0.25cm}   \hspace{0.25cm}
\includegraphics[width=0.19\textwidth]{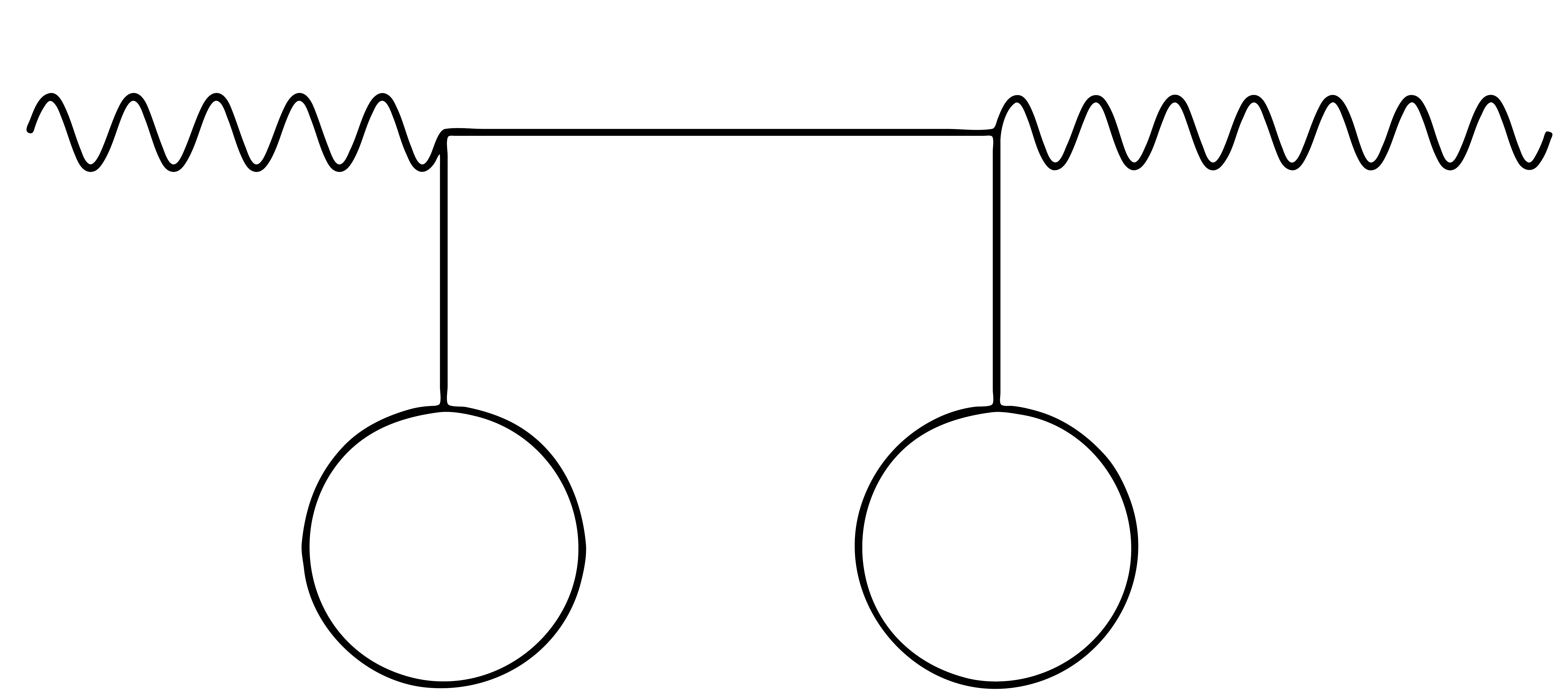}
}
\vspace{-0.3cm}
\caption{Vanishing Non-Gaussian Source due to zero momentum scalar propagator
}
\label{figvanishbyzeromomentum}
\end{figure}

\begin{figure}
\centering{ 
\includegraphics[width=0.22\textwidth]{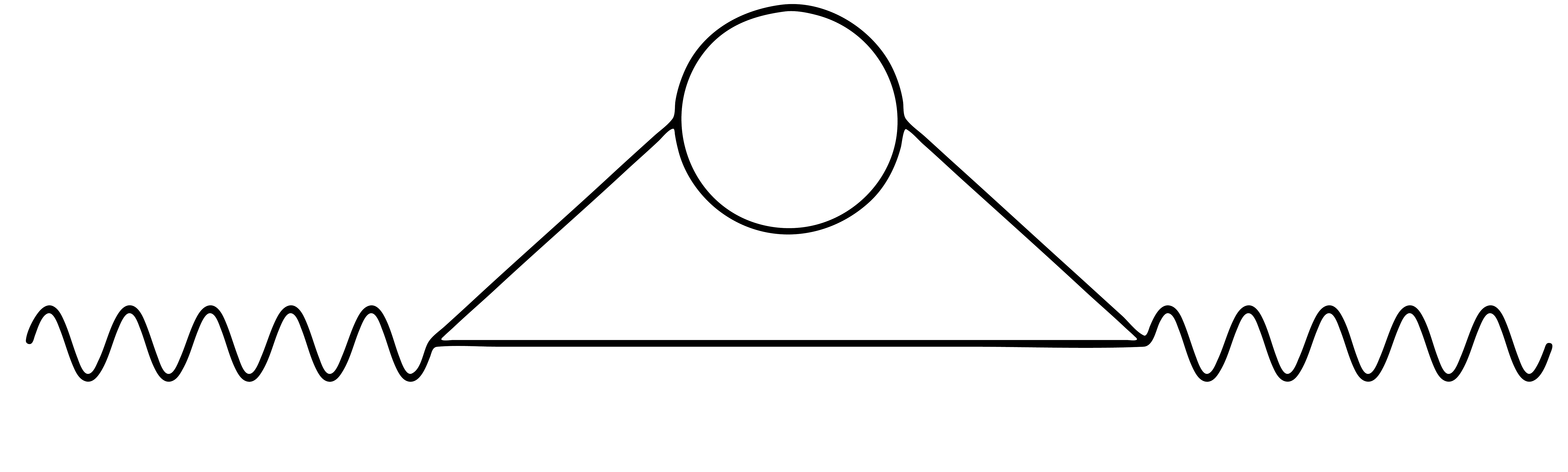}
\hspace{0.3cm}
\includegraphics[width=0.17\textwidth]{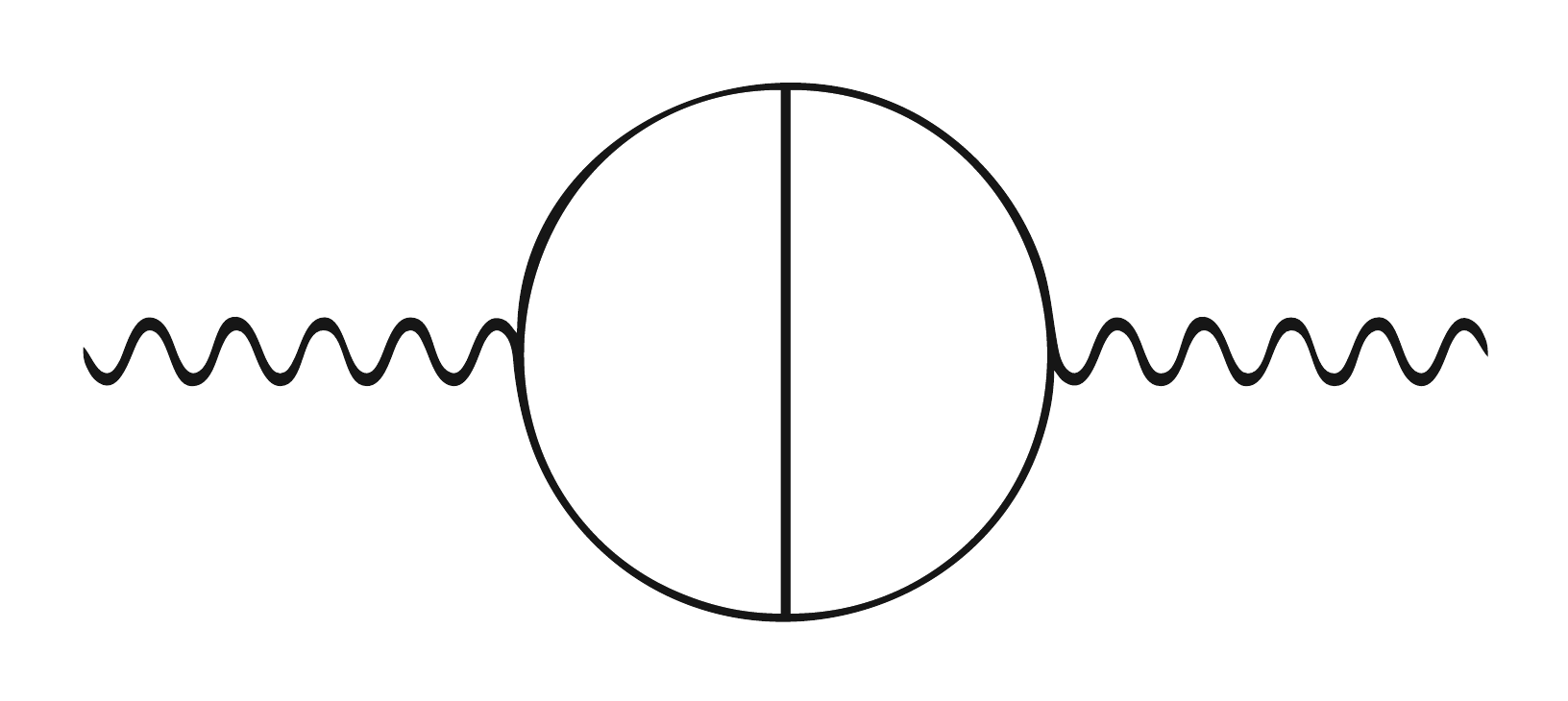}
}
\vspace{-0.3cm}
\caption{Diagrams contributing at  $f_{\rm NL}^2$ Order : Hybrid and Walnut
}
\label{fighybridwalnut}
\end{figure}

\begin{figure}
\centering{ 
\includegraphics[width=0.155\textwidth]{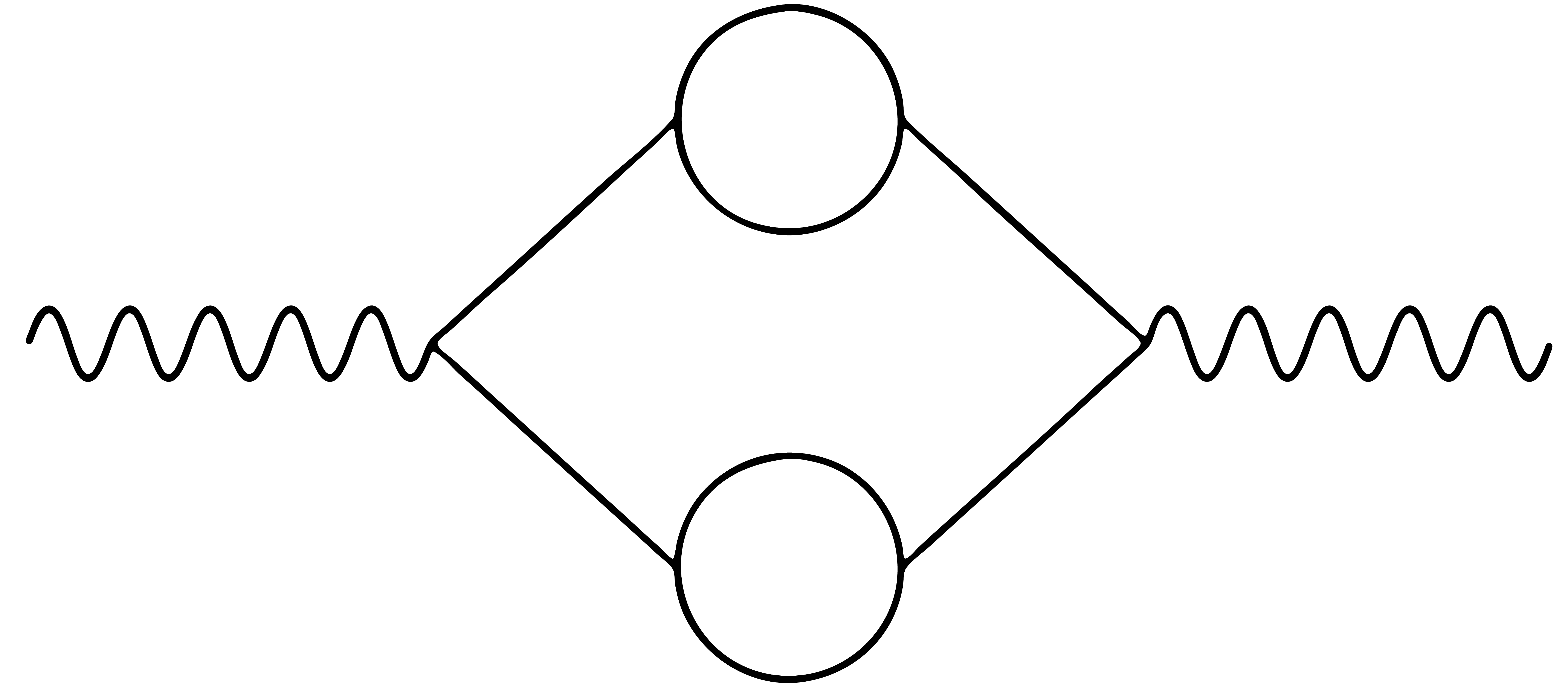}
\includegraphics[width=0.155\textwidth]{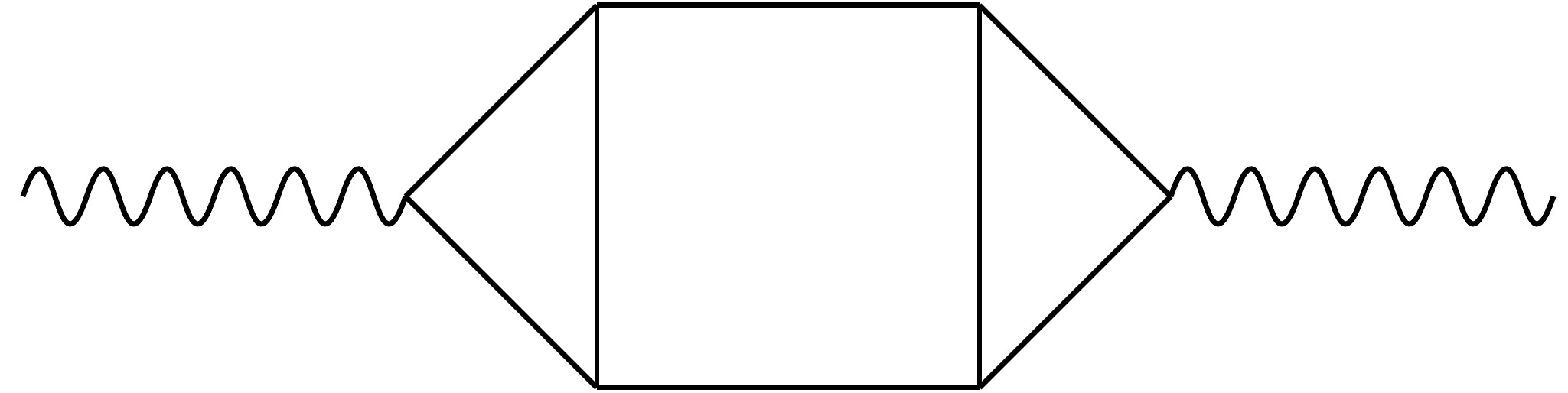}
\includegraphics[width=0.155\textwidth]{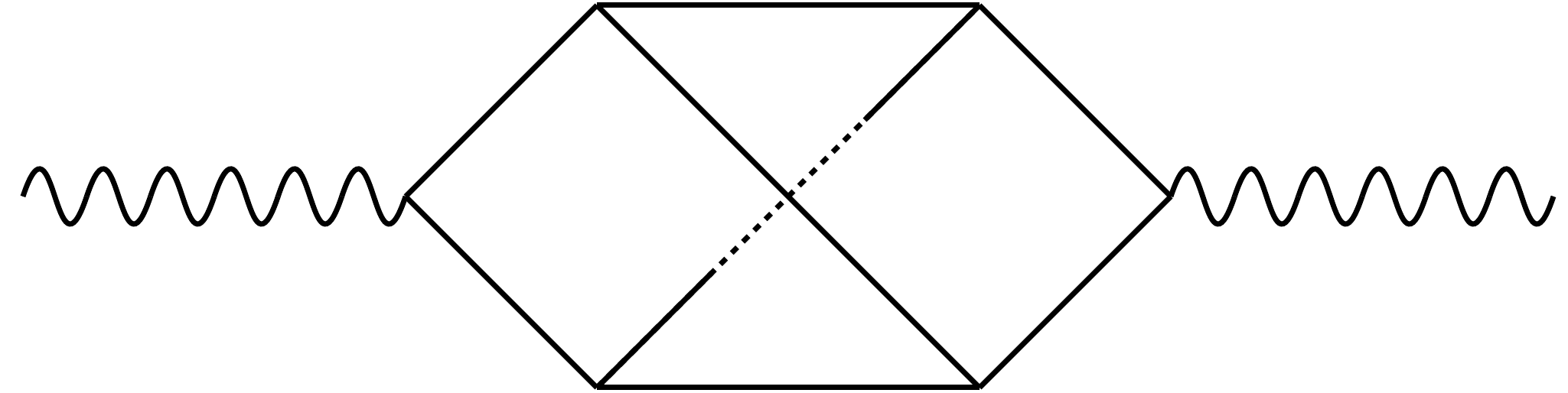}
}
\vspace{-0.3cm}
\caption{Diagrams contributing at $f_{\rm NL}^4$ Order : Reducible, Planar, Non-Planar
}
\label{figfnl4}
\end{figure}

Non-Gaussian scalar fluctuations contribute to the induced GW background at $f_{NL}^2$ and $f_{NL}^4$ order via the 2 loop and 3 loop diagrams, respectively. There are 5 diagram topologies at $f_{NL}^2$ order, all shown in figures \ref{figvanishbysymmetry}, \ref{figvanishbyzeromomentum} and \ref{fighybridwalnut}. Some of these diagrams vanish and some of them contribute to the stochastic GW background. Moreover, there are more diagram topologies at $f_{NL}^4$ order, and we show only the non-vanishing ones in Figure \ref{figfnl4}. Here we discuss all the diagrams in detail.

\emph{ Vanishing Diagrams with the Non-Gaussian Source:}  The three diagram topologies shown in Figures \ref{figvanishbysymmetry} and \ref{figvanishbyzeromomentum} do not contribute to the GW spectrum.  The one in Figure \ref{figvanishbysymmetry} vanishes due to rotational invariance. The two diagrams in Figure \ref{figvanishbyzeromomentum} vanish since the tensor mode couples to scalar modes via derivative couplings. As the momentum  of the propagator vanishes, this vertex does not contribute to the tensor two-point function. The same line of reasoning applies at ${\cal O} (f_{NL}^4)$ to the diagrams with zero-momentum scalar propagators at $h \zeta \zeta$ vertex.

\emph{ Contributions at ${\cal O} (f_{\rm NL}^2)$}:
There are two diagram topologies at $f_{\rm NL}^2$ order that gives non-vanishing contribution to GWs.  One diagram has both tree level and self-corrected scalar propagator, labeled as ``Hybrid" (see the left panel of Figure \ref{fighybridwalnut}). Since the loop correction to $\zeta$ can be factorized, this diagram can be reduced to two one loop calculations. The 
other diagram given on the right panel of Figure \ref{fighybridwalnut} is labeled as ``Walnut" and is a non-reducible two-loop correction.

%

\emph{ Contributions at ${\cal O} (f_{\rm NL}^4)$}:
Neglecting the diagrams vanishing due to zero-momentum scalar propagators at $h \zeta \zeta$ vertex, we have three non-zero diagram topologies at $f_{\rm NL}^4$ order given in Figure \ref{figfnl4}. The left diagram, labeled as ``Reducible", can be expressed as three one loop diagrams (notice the two loops result from the scalar self correction). However, the other two diagrams are 3-loop diagrams which cannot be further factorized. We label these two diagrams as ``Planar"  and ``Non-Planar" due to their topological properties. In this work, we only compute Reducible diagram and estimate the contribution of the other two using earlier studies. These three topologies have been studied in \cite{Garcia-Bellido:2017aan} and it is found that Reducible and Planar diagrams have contributions within ${\cal O}(1)$ proximity and Non-Planar one is suppressed with respect to these two. Hence, in the rest of this work, we will take the total contributions from ${\cal O} (f_{\rm NL}^4)$ diagrams as twice the contribution of the Reducible diagram.


\paragraph{{\bf Results and Discussion.}}
\label{secresults}

We obtain the spectrum of induced GWs by evaluating the two-point function given in  \eqref{eqPhind} at a moment when the corresponding mode is well inside the horizon (ie. $k \eta \gg 1$). The tensor modes are sourced around the horizon crossing, and GWs propagate freely afterwards. Since GWs are relativistic, their energy density scales like radiation, and one can obtain their current energy density using \eqref{eqOGW0}.  

\begin{figure}
\centering{ 
\includegraphics[width=0.43\textwidth]{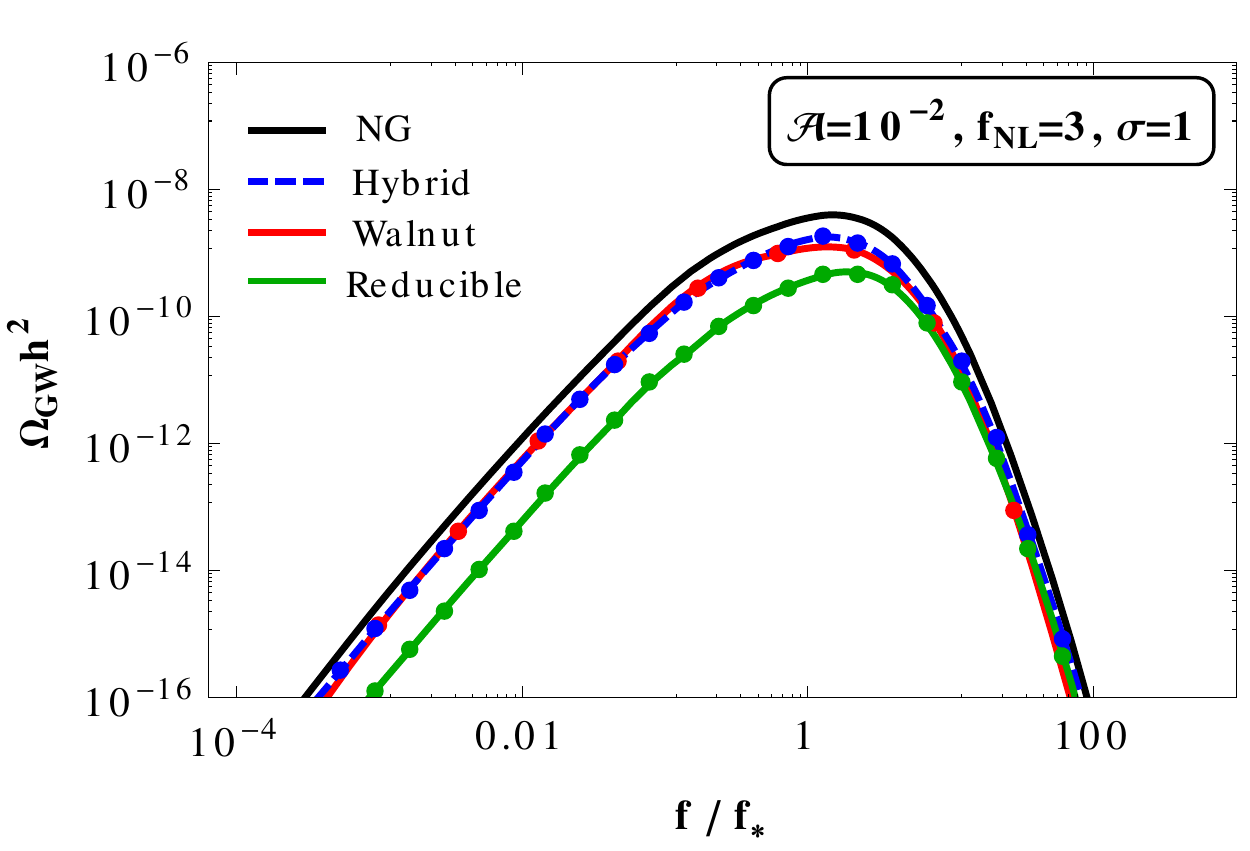}

}
\vspace{-0.3cm}
\caption{Comparison of GW backgrounds from NG diagrams
}
\label{figNGcontributions}
\end{figure}

In Figure \ref{figNGcontributions}, we show the contributions from Hybrid (dashed-blue), Walnut (solid-red) and Reducible (solid-green) diagrams together with total induced GW spectrum (solid-black) originating from non-Gaussian scalar sources (estimating $\Omega^{\rm Reduced}+\Omega^{\rm Planar}+\Omega^{\rm Non-Planar}\simeq2\cdot\Omega^{\rm Reduced}$ as indicated in the previous section). 
The Walnut and Hybrid diagrams give almost same contribution to the GWs at every scale and the Reducible diagram is about an order of magnitude smaller.
%

\begin{figure}
\centering{ 
\includegraphics[width=0.43\textwidth]{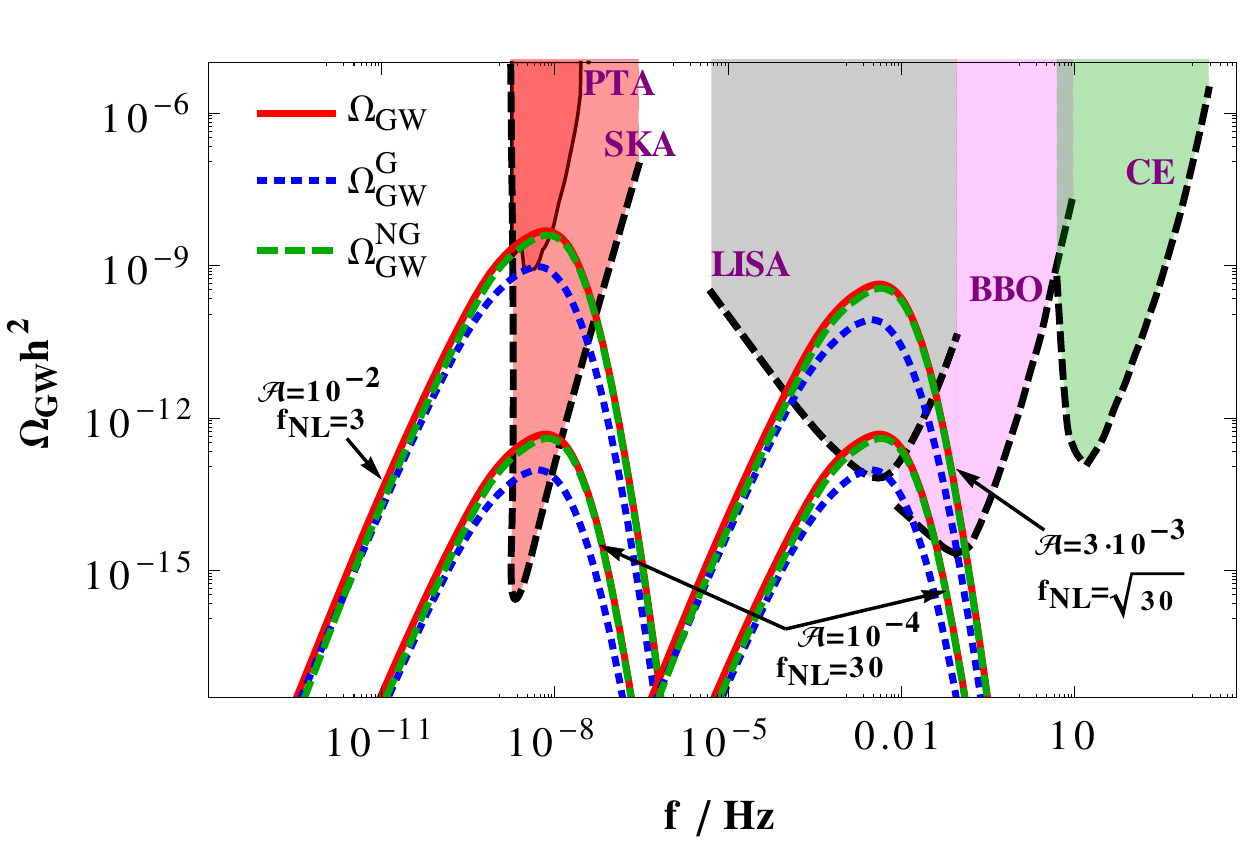}
}
\vspace{-0.3cm}
\caption{$\Omega_{\rm GW}h^2$ at distinct scales from G and NG Sources
}
\label{OGW=OG+ONG}
\end{figure}

In Figure \ref{OGW=OG+ONG}, we show the stochastic GW background induced by Gaussian (dotted-blue) and non-Gaussian (dashed-green) scalar fluctuations together with the total induced GWs (solid-red) at nHz and mHz bands. We set $\sigma=1$ for all curves. The labels $\Omega^G$ and $\Omega^{NG}$,  indicate the ``\emph{source}" of induced GWs (G or NG scalar fluctuations), but not the statistical properties of GWs because whether the source for GWs is Gaussian or not, the resultant GW background is non-Gaussian as it emerges from a cubic interaction  \cite{Espinosa:2018eve,Bartolo:2018evs,Bartolo:2018rku} (see  \cite{Caprini:2018mtu,Bartolo:2018qqn} for a detailed discussion about the stochastic GWs). In all four cases, ${\cal A} \, f_{\rm NL}^2 = 9\cdot 10^{-2}\ll1$ giving $P_\zeta^G > P_\zeta^{NG}$ all wavenumbers. However, although NG part of the scalar fluctuations are subdominant to G part, the GWs produced from  NG scalar fluctuations are about an order of magnitude larger, $\Omega^{NG}_{GW} \simeq 9 \, \Omega^G_{GW}$.

For both LISA and PTA scales, there are two example GW signals.
The larger amplitude GWs originate from scalar fluctuations leading to abundant amount of PBHs, which constitute the totality or considerable fraction of the DM  \footnote{ \label{betafootnote} In the presence of positive non-Gaussianity parameter, $\beta$ can be expressed as $\beta = \, {\rm erfc} \left( Y_+(\zeta_{\rm th}) \right) + \, {\rm erfc} \left( Y_-(\zeta_{\rm th}) \right)$, where  $Y_\pm(\zeta) = \frac{1}{2 \sqrt{2  {\cal A}} \, f_{\rm NL} \,} \left( -1 \pm \sqrt{ 1+ 4 f_{\rm NL} (f_{\rm NL } \, {\cal A} +\zeta)} \right)$  \cite{Byrnes:2012yx}. The formation fraction, $\beta$, can be connected to the ratio of the energy density of PBH  to DM in the current universe  ${\cal F}(M) \equiv \frac{1}{\rho_{\rm DM}}\frac{d \rho_{\rm PBH}}{ d\ln M} \simeq 6.7 \cdot 10^8 \, \gamma^{1/2} \, \beta(M) \sqrt{\frac{M_\odot}{M}} $ \cite{Khlopov:2008qy,Carr:2016drx},  where $M_\odot$ is the solar mass.
}  ( $\zeta_{\rm th} \simeq $ 1.4).  The primordial scalar fluctuations resulting in PBH DM have different amplitudes and $\beta$ values for different ($\beta_{\rm LISA} \neq \beta_{\rm PTA}$)  scales. This is because different mass PBHs form at different times, hence their energy density grows relative to the background energy density different amount (see footnote \ref{betafootnote}). 
On the other hand, the small amplitude signal is chosen near the sensitivity limit of the both GW experiments to show the large range of scalar fluctuations producing visible signals. Although we illustrate the signatures only at PTA \cite{Arzoumanian:2015liz,Lentati:2015qwp,Shannon:2015ect}, SKA \cite{Moore:2014lga,Zhao:2013bba} and LISA \cite{AmaroSeoane:2012km,Bartolo:2016ami,Audley:2017drz} scales, this study can be extended for other GW experiments such as DECIGO/BBO \cite{Yagi:2011wg} and Cosmic Explorer (CE) \cite{Evans:2016mbw}.

\begin{figure}
\centering{ 
\includegraphics[width=0.4\textwidth]{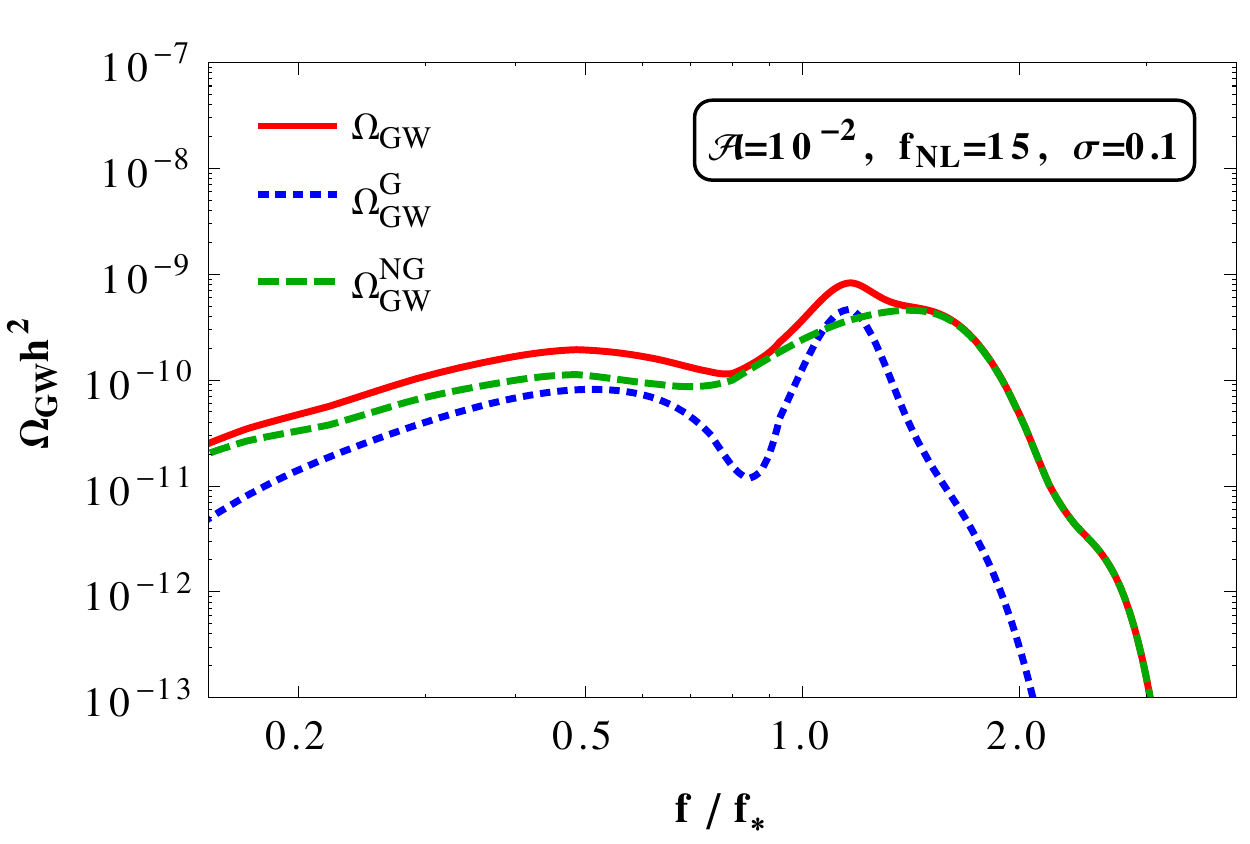}
\includegraphics[width=0.4\textwidth]{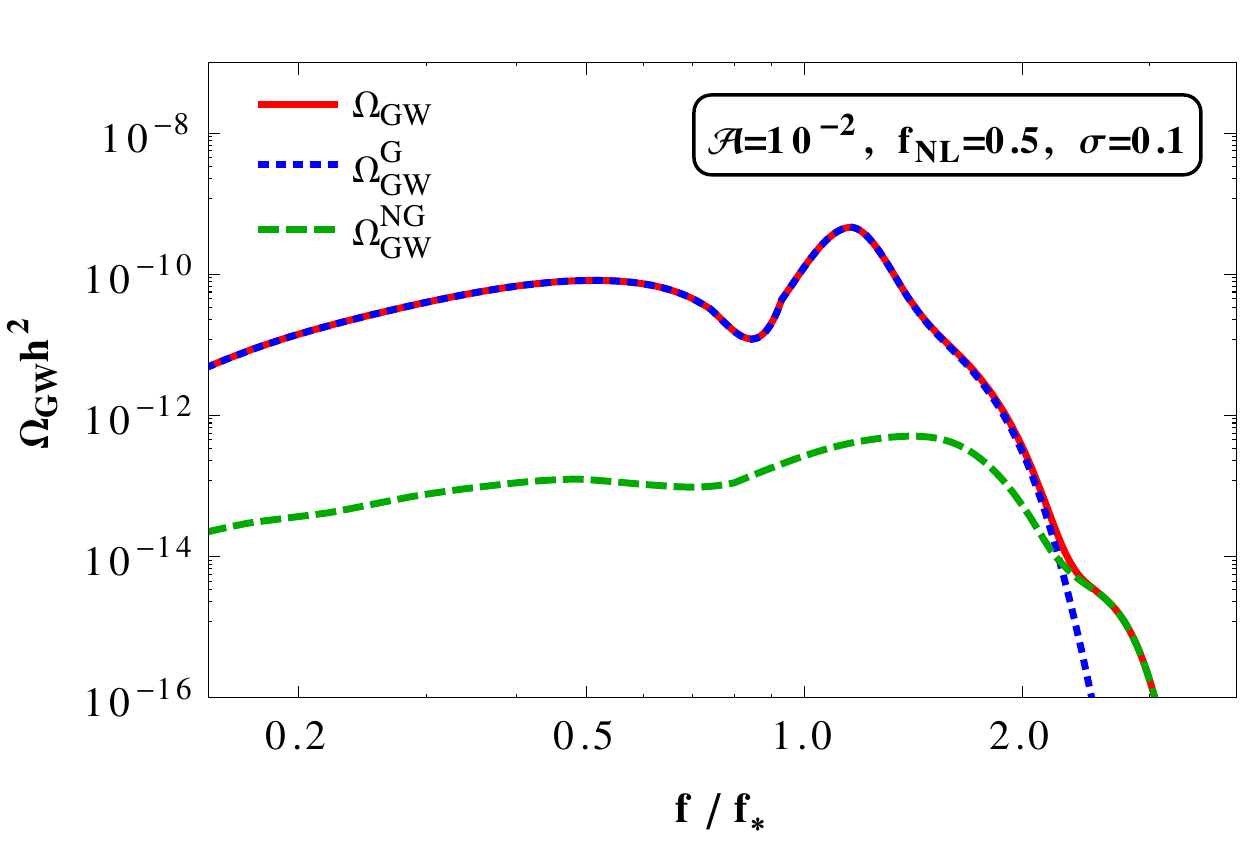}
}
\vspace{-4mm}
\caption{Observational signatures of local NG in the UV regime of the GW spectrum
}
\label{fig:UVsignaturesofNG}
\end{figure}

NG scalar fluctuations peak at a larger wavenumber with respect to G fluctuations, and they have larger width since NG fluctuations arise from the convolution of G fluctuations (see \eqref{defnG-NG} and Figure \ref{figPZ}). As a result of this, the Reducible diagram peaks at the highest frequency, the Hybrid and Walnut diagrams follow it and the Gaussian diagram has a peak at the lowest frequency (see Figures \ref{figNGcontributions} and  \ref{OGW=OG+ONG}). As they peak at larger wavenumber, NG density fluctuations can easily dominate the UV tail of the induced GW spectrum, and this can lead to observational signatures as shown in Figure \ref{fig:UVsignaturesofNG}. These signatures show themselves prominently for small width scalar signals, but the modification in the UV tail also exists for large width scalar signals with visually less differentiable way. Therefore, in order to show this effect boldly, we choose a narrow width signal, $\sigma=0.1$. One example is the top panel in Figure \ref{fig:UVsignaturesofNG}, where $f_{\rm NL}=15$ and GWs from NG and G sources have comparable amplitude. In that panel, in addition to the bump around $f\simeq 1.2f_*$, there exist a sharp slope change/second bump in the large frequency part of the spectrum due to the peak of the induced GWs from NG fluctuations around $f\simeq 1.7f_*$. Another concrete example is in the bottom panel of Figure \ref{fig:UVsignaturesofNG}, for $f_{\rm NL}=0.5$. Although, the peak of the induced GWs from NG perturbations are about 2 orders of magnitude smaller with respect to GWs sourced by G fluctuations, since the GWs from NG have a longer UV tail, this still allows local non-Gaussianity to leave its imprint around $f\simeq 2f_*$. Hence, if the stochastic GW background is detected, but there is no sign of primordial non-Gaussianity, this constrains the NG parameter as  ${\cal A} \cdot f_{\rm NL}^2 \la 10^{-2}$ for the experiments reaching sensitivity of $\Omega_{\rm GW} h^2\sim 10^{-15}$. This indicates that sensitive next-generation-experiments such as  PTA-SKA, LISA, DECIGO, BBO, CE and ET  may probe NG parameter as low as $f_{NL}\sim 0.5$ which is even better than estimated sensitivity of the next generation CMB experiments. With higher sensitivity GW experiments, the primordial local NG can be constrained even more stringently via its mark on the UV part of the spectrum.


In the IR regime ($k\ll k_*$), we have $k_* \sim p \sim \vert \bk - \bp \vert  \gg k$ as the support of the momentum integrals are around the wavenumber $k_*$, hence $F_T(u,u) \simeq \frac{27}{u^2} \sin^2(u / {\sqrt 3})$ where $u \equiv p/k \gg 1$. In result, $P_{h_i}\propto {k^3} \int \, \frac{F_T^2}{k^4} d  (k \eta') (k \eta') d (k \eta'') (k \eta'') \propto k^3$. Therefore, in the IR regime,  induced GWs scale with (nearly) the third power of the wavenumber independent of the fact that they are sourced by G or NG fluctuations.

\paragraph{ {\bf Large ``$f_{\rm NL}$" ( $\chi^2$ ) Limit.}}
\label{sec:largefnl}

In the limit $f_{NL}^2 {\cal A} \gg 1$ (we stress that  the expression in \eqref{eqnscalarfluct}  is not considered as a perturbative series expansion in orders of $f_{\rm NL}$), we can absorbe the $f_{\rm NL}$ parameter into definition of $\zeta_G$ such that $\zeta \simeq \pm( \zeta_G^2 - \langle \zeta_G^2 \rangle )$; hence the density fluctuations obey the $\chi^2$ statistics. The relevant diagrams in that case are given in Figure \ref{figfnl4}. The PBH production and induced GW spectrum from $\chi^2$ statistics have been studied for bumpy scalar fluctuations in \cite{Garcia-Bellido:2017aan}. (See \cite{Byrnes:2012yx,Young:2013oia} for a detailed discussion on the effects of non-Gaussianity on PBH production, also \cite{Lyth:2012yp}, and \cite{Nakama:2016gzw} for an estimate of the GW background from non-Gaussian fluctuations.)

The induced GW background is proportional to even powers of $f_{\rm NL}$, so the sign of the $f_{\rm NL}$ does not change the tensor two-point function. However, since the probability distribution function of the curvature fluctuations gets skewed, the PBH production efficiency increases (decreases) for $f_{NL}>0$ ($f_{NL}<0$) \cite{Byrnes:2012yx,Young:2013oia}. Therefore with $f_{\rm NL}>0$, the same amount of PBH can be produced by smaller amplitude (variance) scalar fluctuations, which results in smaller amplitude induced GWs \cite{Nakama:2016gzw,Garcia-Bellido:2017aan}.



\paragraph{{\bf Primordial Black Holes.}}
\label{secpbh}

When a mode with an amplitude larger than the collapse threshold re-enters the horizon, the energy content inside the horizon collapses to a PBH that have a mass of the order of the mass inside the horizon. Since the mass of the PBH is comparable to the horizon mass, the PBH mass can be related to the wavenumber of the mode as $M_{\rm PBH} =   20 \, \gamma \, M_\odot \left( k/ {\rm pc}^{-1} \right)^{-2} $ or to its frequency   \cite{Garcia-Bellido:2017aan} 

\vspace{-5mm}

\begin{equation}
M_{\rm PBH} =  50 \, \gamma \, M_\odot \left( \frac{10^{-9} \, {\rm Hz}}{f}  \right)^2 \, , 
\label{Mpbhf}
\end{equation}
where $\gamma$ indicates the portion of the horizon that becomes PBH (we assume $\gamma=0.2$ throughout this work). For a given scale, the fraction of the universe that turns into a PBH can be expressed as
\begin{equation}
\beta(k) = \int_{\zeta_{\rm th}}^\infty {\cal P}(\zeta_{\rm cg} (k)) \,\, d\zeta_{\rm cg} (k) \; ,
\end{equation}
where ${\cal P}$ is the probability distribution function (p.d.f) of the curvature fluctuations, $\zeta_{\rm th}$ is the threshold value ${\cal O}(0.1-1)$, and $\zeta_{\rm cg}(k)$ is the coarse-grained density contrast : $\zeta_{\rm cg}(k) = \int W(k)  \zeta(k) \, d\ln k$, where $W$ is a smoothing window function. Because PBHs are produced from the tail of the p.d.f of the curvature perturbations, the PBH amount is extremely sensitive to both threshold value and the details of the distribution function. Therefore, even small differences in either of them  can lead to many orders of magnitude changes in PBH amount. 

It is a remarkable possibility that PBHs can be whole or considerable fraction of DM \cite{Bird:2016dcv,Clesse:2016vqa,Sasaki:2016jop}. The two mass ranges, $M \sim10 M_\odot$ and $M \sim 10^{-12} M_\odot$, are being discussed for such possiblity \cite{Carr:2016drx,Zumalacarregui:2017qqd,Garcia-Bellido:2017imq}. It is even further interesting that the modes which produce the PBHs in these mass ranges lead to inevitable induced GW backgrounds. These backgrounds correspond to the frequency bands of nHz and mHz \eqref{Mpbhf} which will be probed by PTA, SKA and LISA experiments with enormous precision in close future \cite{Inomata:2016rbd,Garcia-Bellido:2016dkw,Garcia-Bellido:2017aan,Barack:2018yly}. However, whether PBHs are the dominant or completely negligible portion of the DM,  induced GWs can reveal valuable information about the small scales.

\paragraph{{\bf Summary and Conclusions.}}
\label{sec:conclusion}

We show that the primordial quantum fluctuations can be probed via the stochastic GW background  induced by the density perturbations re-entering the horizon during radiation dominated era. Although subdominant, the non-Gaussian component of the density fluctuations can produce more GWs than the Gaussian component. Furthermore, the GWs sourced by NG scalar fluctuations peaks at a larger frequency which might result in distinctive signatures.  From the UV part of the induced GW spectrum, the sensitive next-generation-experiments such as  PTA-SKA, LISA, DECIGO, BBO, CE and ET  can detect (or constrain) NG parameter as low as $f_{NL}\sim 0.5$, which is even better than predictions of the next generation CMB experiments (see Figure \ref{fig:UVsignaturesofNG} and relevant discussion). If induced GW background is not detected, this translates into bounds on the amplitude of scalar fluctuations which is orders of magnitude stronger than the bounds on scalar fluctuations due to PBHs amount.

With positive $f_{\rm NL}$,  scalar fluctuations can efficiently produce PBHs which might constitute the whole or some fraction of DM. Two mass ranges for this intriguing possibility are $M \sim10 M_\odot$ and $M \sim 10^{-12} M_\odot$, and these PBHs are accompanied by induced GWs in nHz and mHz bands respectively, which will be probed by PTA-SKA and LISA with great precision. However, whether PBHs are DM or not, primordial non-Gaussianity in scalar sector can be detected via induced GWs for a wide range of amplitudes.


\vspace{3mm}
\paragraph{ {\bf Acknowledgments.}}
It is a pleasure to thank Nicola Bartolo,  Juan Garcia-Bellido, Marco Peloso, Ignacy Sawicki, Lorenzo Sorbo, Bayram Tekin and Vincent Vennin for their critical reading of the draft, their feedback and useful discussions. We also thank Sohyun Park, Lorenzo Reverberi and Tom Zlosnik for illuminating discussions.  The author acknowledges the encouragement and support of \"Omer and Hamiyet \"Unal during this project. Part of this research project was conducted using computational resources at Physics Institute of the CAS and University of Minnesota, and we acknowledge the help of Josef  Dvoracek and Alexandros Papageorgiou for this process.  This work is supported by European Structural and Investment Funds and the Czech Ministry of Education, Youth and Sports (Project CoGraDS - CZ.$02.1.01/0.0/0.0/15\_003/0000437$).

\vspace{3mm}

\paragraph{ {\bf Note Added:}} As this work is being completed, Ref. \cite{Cai:2018dig} appeared which also studies the effects of NG scalar fluctuations on GW spectrum. However, their investigation focuses on the large $f_{\rm NL}$ limit, $\chi^2$ distribution, hence their scalar fluctuations are dominantly Non-Gaussian and different than this study. Ref. \cite{Cai:2018dig} calculates the GW spectrum using the first diagram given in Figure \ref{figfnl4}, neglecting the other two diagrams. We think this gives gives the correct order of magnitude, but not precise result. We note that the large $f_{\rm NL}$ limit, or $\chi^2$ distribution, had been already studied in Ref. \cite{Garcia-Bellido:2017aan} with all the diagrams in Figure \ref{figfnl4}.

\end{document}